**Scattering-force-driven light-sheet fluorescence microscopy for scan-free three-dimensional imaging**

*Sen Geng[1], Zhuoren Wan[1,2], Zhaoyang Wen[1,2], Dongxu Zhu[1], Yuan Chen[1], Wending Huang[1], Zijian Wang[1,2*], Ming Yan[1,2,3*], and Heping Zeng[1,2,4]*

[1]State Key Laboratory of Precision Spectroscopy, and Hainan Institute, East China Normal University, Shanghai 200062, China

[2]Chongqing Key Laboratory of Precision Optics, Chongqing Institute of East China Normal University, Chongqing 401120, China

[3]Collaborative Innovation Center of Extreme Optics, Shanxi University, Taiyuan, Shanxi 030006, People's Republic of China

[4]Jinan Institute of Quantum Technology, Jinan, Shandong 250101, China

E-mail: zjwang@lps.ecnu.edu.cn; myan@lps.ecnu.edu.cn

**Abstract:** Optical manipulation offers a powerful non-contact approach to precise positioning, directional transport and three-dimensional (3D) imaging of microscale objects. However, positional fluctuations during optical confinement introduce spatial sampling errors that compromise imaging quality and acquisition speed. Here, we demonstrate a scattering-force-driven optical manipulation imaging method that replaces the stepwise scanning of conventional optical trapping with continuous directional transport of microparticles. An axial scattering force generated by a low-numerical-aperture focused beam propels microparticles through a light sheet, while their positions and fluorescence signals are simultaneously recorded to establish direct spatial correspondence between particle motion and tomographic images. Experiments demonstrate that this approach reduces the effects of positional fluctuations on 3D imaging, achieving an axial sampling interval of 0.1 μm and an imaging rate of 250 Hz. The method also enables effective imaging of samples with diverse structural features while reducing optical power requirements by 70%. By combining robust spatial sampling, high-speed acquisition and reduced photodamage, this approach offers a promising route to rapid 3D imaging, dynamic observation and microscale spatial analysis of microparticles and biological specimens.

## 1. Introduction

Optical manipulation exploits momentum exchange between light and matter to generate optical forces, enabling controlled regulation of the position, direction of motion, and trajectory of microparticles [1-4]. When integrated with microscopy, optical manipulation can actively regulate the relative position between a target and the imaging region, guiding the target through the detection volume along a prescribed trajectory [5,6]. This capability overcomes the limitations of fixed fields of view and passive observation, providing a new spatial sampling approach for dynamic imaging and three-dimensional (3D) information acquisition [7,8]. However, optical manipulation is susceptible to spatial fluctuations in the scattering and gradient force distributions, environmental perturbations, and Brownian motion [9,10], which can introduce random deviations in the actual particle position and trajectory [11-13]. When these positional fluctuations directly affect spatial sampling, discrepancies arise between the assigned sampling coordinates and the actual particle positions [14], resulting in image misregistration and spatial information distortion and ultimately limiting the axial sampling and reconstruction quality of 3D imaging [15-17].

To mitigate the influence of positional fluctuations on optical-manipulation-assisted imaging, various stabilization and correction strategies have been developed [18-20]. One approach improves position-detection sensitivity to track particle motion in real time and compensates for positional deviations through feedback control or model-based correction, thereby reducing fluctuation-induced measurement errors [11]. Another approach strengthens spatial confinement using optical traps to suppress random displacements caused by Brownian motion and environmental perturbations [21-23]. More recently, holographic optical tweezers (HOT) have been incorporated into tomographic imaging [24-26], where enhanced 3D optical confinement restricts positional fluctuations of suspended samples to the tens-of-nanometers range, enabling tomographic imaging with an axial step size of 0.16 μm at an acquisition rate of 10 Hz [14]. In such optical-manipulation-assisted imaging systems, the axial scanning step and imaging speed are closely related to the manipulation scheme [27-29], because positional fluctuations of the confined particle directly affect spatial sampling and tomographic reconstruction [30-32]. Nevertheless, stronger optical confinement mainly improves stability by restricting the range of particle motion and cannot eliminate discrepancies between the actual and prescribed positions during dynamic manipulation, leaving uncertainty in the spatial sampling coordinates [33-35]. Moreover, sustained optical confinement and feedback control increase system complexity, resulting in a trade-off between imaging speed and low photodamage [36,37].

To address these limitations, we propose a scattering-force-driven active optical manipulation technique for 3D imaging. Unlike conventional optical tweezers, which rely on high-numerical-aperture (NA) focusing to generate strong gradient forces for stable trapping, our approach employs a low-NA objective to weaken the axial gradient restoring force, allowing the forward scattering force to dominate the axial force balance. The resulting momentum transfer drives microparticles to move steadily along the direction of light propagation without sustained optical trapping. By eliminating the need for complex optical fields to maintain motion stability, this strategy requires neither a spatial light modulator (SLM) nor a mechanical translation stage, thereby reducing system complexity and mitigating the influence of trap fluctuations on spatial information acquisition. Continuous particle motion reduces the axial sampling interval to 0.1 μm and increases the image acquisition rate to 250 Hz. Furthermore, the required optical power can be as low as 15 mW, potentially reducing photothermal effects and photodamage associated with prolonged illumination and providing a new optical manipulation approach for rapid, low-photodamage 3D imaging with fine axial sampling.

## 2. Methods

### 2.1. Basic concept

In this study, we propose an active optical manipulation imaging method based on optical scattering forces. Rather than using gradient-force-dominated optical traps to hold microparticles at fixed positions, the method exploits optical scattering forces to propel them through the imaging region. **Figure 1(a)** outlines the optical arrangement: an upper objective focuses the driving beam, a lateral light-sheet module excites successive sections of the sample, and a lower high-NA objective collects fluorescence and scattered driving light.

**Figure 1(b)** illustrates the spatial alignment of the focal regions of the three objectives and the corresponding fluorescence excitation and optical-force conditions of a microparticle within the coupled optical fields. The thinnest region of the light sheet, the center of the driving optical field, and the focal plane of the imaging objective are spatially co-aligned. Within the driving field, the microparticle is primarily subjected to two optical forces: the scattering force ($F_{scat}$), directed along the direction of light propagation, and the axial gradient force ($F_{grad,z}$), directed toward the center of the focused optical field.

For a focused beam propagating along the z direction, the resultant axial optical force acting on the microparticle can be expressed as [38]:

$$F_z = F_{scat} - F_{grad,z} \tag{1}$$

Here $F_z$ denotes the axial force along the direction of light propagation, and z denotes the axial position of the microsphere. The resulting particle state can be determined according to the following criterion:

$$\begin{cases} F_z(z_0) = 0, \left.\frac{\partial F_z}{\partial z}\right|_{z_0} < 0 & \text{Trapping} \\ F_z(\mathrm{z}) > 0 & \text{Propulsion} \end{cases} \tag{2}$$

The motion state of the microparticle is governed by the competition between the axial gradient force and the scattering force. The numerical aperture of the objective determines their relative contributions by controlling the beam convergence angle and the intensity gradient near the focus. High-NA focusing generates a strong axial gradient restoring force that can balance the scattering force and establish a stable trapping position. As the NA decreases, the beam convergence angle becomes smaller and the axial gradient restoring force weakens. Once it is insufficient to counteract the forward scattering force, the stable axial equilibrium position disappears, and the microparticle transitions from stable trapping to directional propulsion along the direction of light propagation, as shown in **Figure 1(c)**. Consequently, the dominant axial scattering force maintains continuous directional motion of the microparticle, substantially reducing the relative influence of random displacements induced by Brownian motion and environmental perturbations on axial scanning, while the transverse gradient force confines the microparticle near the optical axis and suppresses lateral random displacement. More importantly, this propulsion mode does not rely on the equilibrium position of a stable optical trap to define the sampling coordinates, thereby avoiding the influence of equilibrium-position variations caused by fluctuations in the optical-force distribution on spatial positioning.

During active manipulation, fluorescent microparticles are continuously propelled by the scattering force through the light sheet for active axial scanning. As a microparticle enters the light-sheet illumination region, its successive axial sections are excited, and the emitted fluorescence is collected by a high-NA objective [39]. **Figures 1(d)** and **1(e)** depict the light-sheet fluorescence scanning process from the X/Z and X/Y views, respectively. In the XZ view, the microparticle travels downward along the optical axis and successively crosses the light sheet; the XY view captures fluorescence cross-sections acquired at successive axial positions, enabling continuous spatial sampling.

This approach transforms the scattering force, conventionally regarded as an unfavorable factor in optical trapping, into the driving force for active scanning, extending optical manipulation from stable trapping to continuous propulsion. Because spatial sampling no longer relies on prescribed optical-trap positions [14], the actual particle position measured for each fluorescence image can be used to determine its spatial coordinate, thereby reducing the influence of discrepancies between actual and prescribed positions on 3D reconstruction. Meanwhile, eliminating the need for sustained strong optical confinement reduces the required optical power, providing a new approach for rapid, low-photodamage 3D imaging of microscale samples [40].

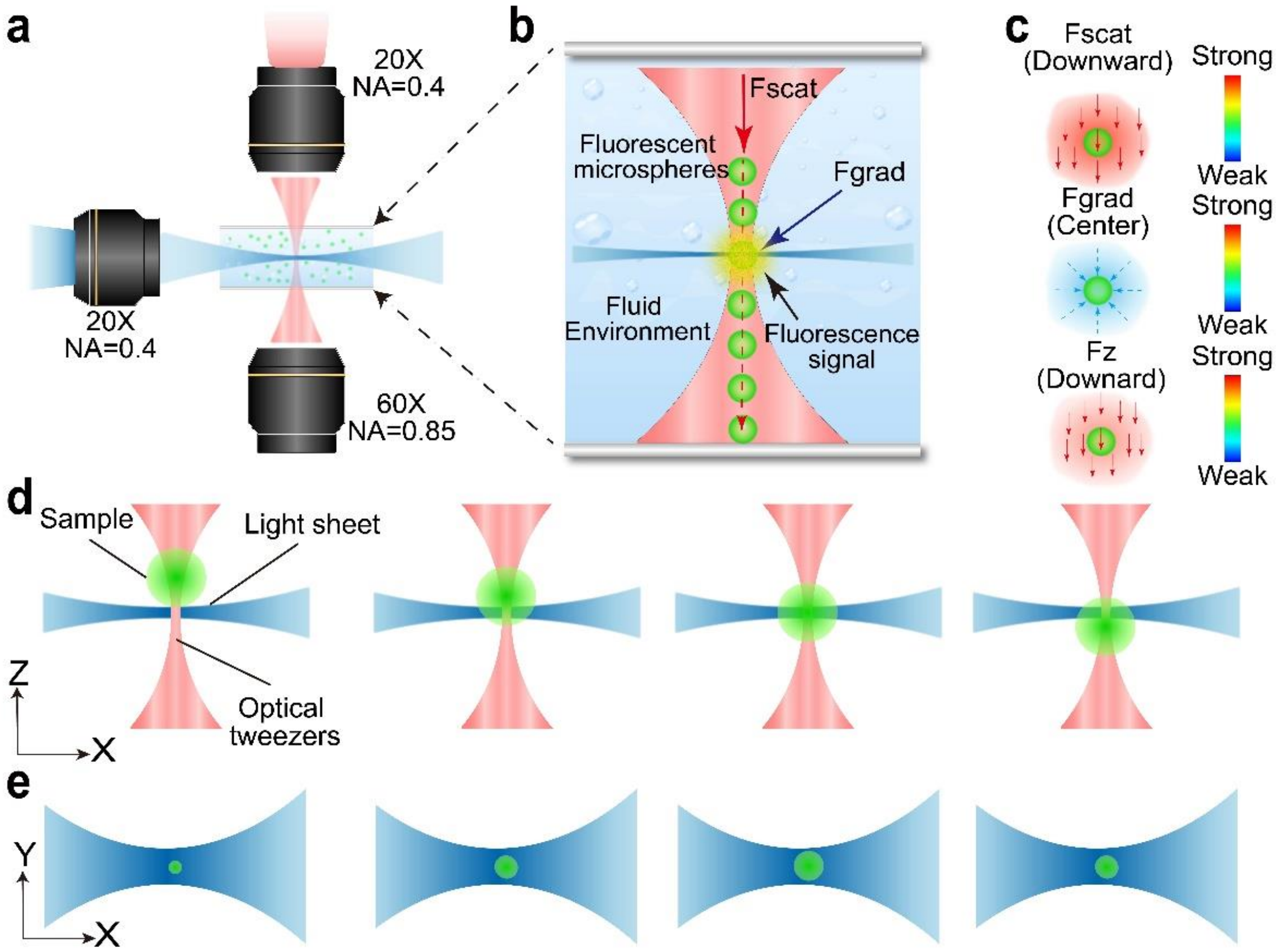

**Figure 1: Schematic of the imaging principle. a,** Principle of scattering-force-driven 3D light-sheet fluorescence imaging without mechanical scanning. **b,** Optical forces acting on a microsphere in the optical field. **c,** When the scattering force exceeds the gradient force ($F_{scat} > F_{grad}$), the axial component of the resultant force ($F_z$) acts along the direction of light propagation, while the transverse component directs the microsphere toward the optical axis. **d** and **e,** Fluorescence excitation regions at successive microsphere positions as the particle passes through the light-sheet center, viewed in the XZ and XY planes, respectively.

### 2.2. Experimental setup

**Figure 2** summarizes the experimental setup, which comprises three modules: optical propulsion, light-sheet fluorescence imaging, and position detection. The driving beam is focused through a low-NA objective above the sample to generate an optical field dominated by the axial scattering force, propelling the microparticle downward along the direction of light propagation. The lateral excitation beam is focused through an objective to form a light sheet for sectional excitation of fluorescent polystyrene microparticles passing through the illumination region (Supplementary Figure 1). A high-NA objective below the sample simultaneously collects the green fluorescence and scattered light from the driving beam for light-sheet fluorescence imaging and particle position detection, respectively. The samples are dispersed in a liquid medium and contained within a square glass capillary.

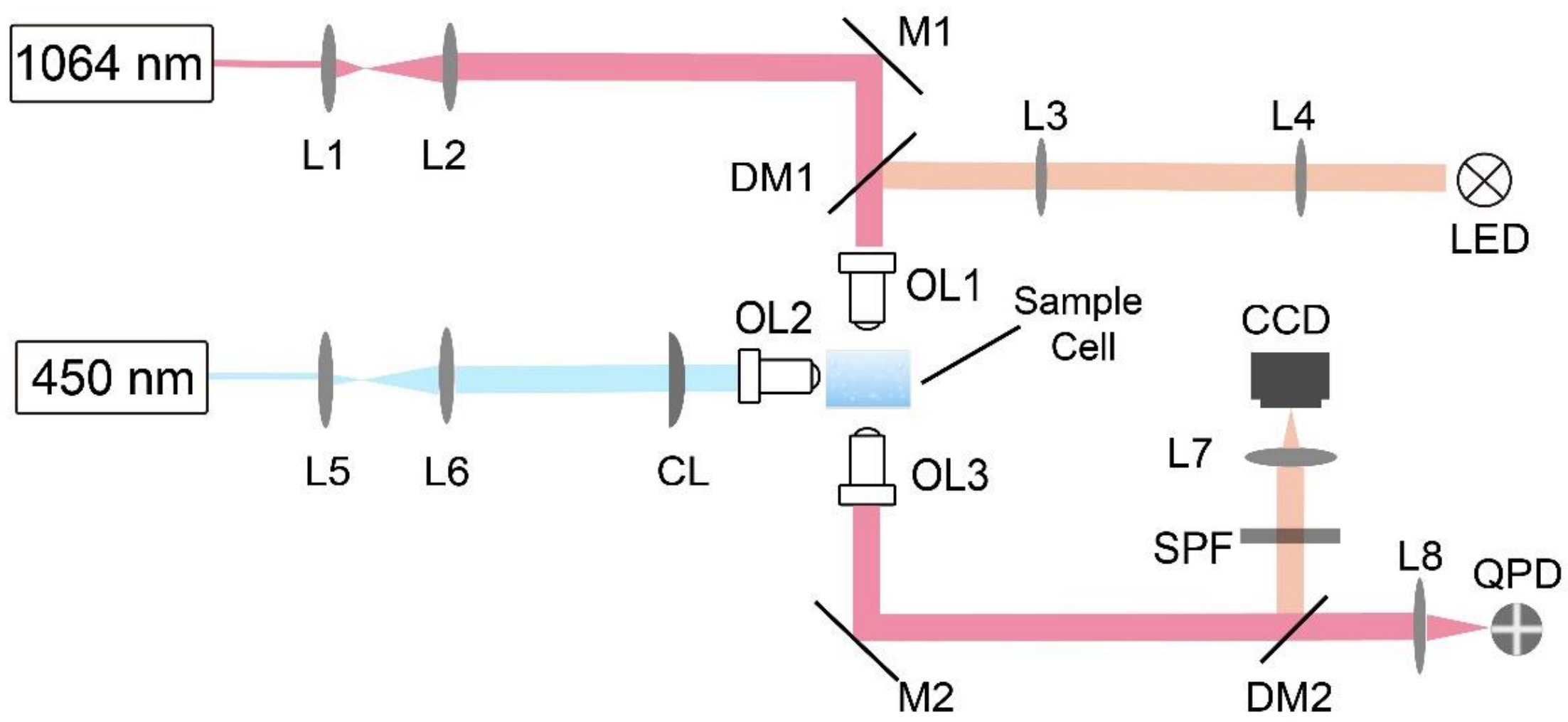

**Figure 2: Schematic of the experimental setup.** The manipulation beam was provided by a 1064 nm continuous-wave diode laser, and fluorescence excitation was provided by a 450 nm continuous-wave diode laser. Optical components: L1–L8, lenses; LED, collimated light source used with L3 and L4 for Köhler illumination; sample chamber, a square capillary with a 3 mm inner width; M, mirror; DM1 and DM2, long-pass dichroic mirrors (cut-on wavelength: 1000 nm); OL1 and OL2, low-NA objectives; OL3, high-NA objective; CL, cylindrical lens (f = 100 mm); SPF, short-pass filter (cutoff wavelength: 500 nm); QPD, quadrant photodiode. The focal lengths of L1, L3, and L5 are 25 mm, those of L2, L4, and L6 are 75 mm, and those of L7 and L8 are 50 mm.

## 3. Results

### 3.1. Theoretical simulation

A laser beam focused by a low-NA objective produces an optical field that exerts a driving force on 15-μm-diameter fluorescent polystyrene microspheres. To investigate the force and motion characteristics of microparticles in a low-NA focused optical field, we established a theoretical model and numerically simulated their motion under the driving beam [41].

Within the sample chamber, the low-NA objective produces an axially propagating Gaussian optical field. Its intensity distribution is given by:

$$I(r,z) = \left[2P/\left(\pi\omega^2(z)\right)\right]exp[-2r^2/\omega^2(z)] \tag{3}$$

In Equation (3), P denotes the laser power at the sample position, r denotes the radial distance between the sample center and the optical axis, and $\omega(z)$ denotes the beam radius at the axial position z.

$$\omega(z) = w_0\left\{1 + \left[\left(z - z_f\right)/z_R\right]^2\right\}^{1/2} \tag{4}$$

In Equation (4), $w_0$ denotes the beam waist radius, $z_f$ denotes the focal position, and $z_R$ denotes the Rayleigh range. The microsphere experiences both axial and transverse optical forces in the optical field, and the total optical force is expressed as:

$$F_{opt} = F_z(r,z)e_z + F_r(r,z)e_r \tag{5}$$

Here, $F_r$ denotes the radial component in Equation (5). The axial optical force is expressed in terms of the optical-force efficiency, and $e_z$ and $e_r$ are unit vectors used to indicate the directions of the two force components:

$$F_z(r,P) = (n_m P/c)Q_z(r.z) \tag{6}$$

$$F_r(r,P) = (n_m P/c)Q_r(r,z) \tag{7}$$

In Equations (6) and (7), $n_m$ is the refractive index of the medium, $c$ is the speed of light in vacuum, and $Q_z$ is the axial optical-force efficiency. The axial gradient restoring force is weakened by using an objective with a numerical aperture of 0.4. Under the laser power, sample size, and effective scanning range used in this study, $F_z(0,z) > 0$, indicating that the microsphere does not have a stable axial trapping position but instead moves along the direction of light propagation under scattering-force-dominated propulsion. At the same time, the transverse optical-intensity gradient exerts radial confinement on the microsphere, ensuring that its trajectory remains within the central region of the optical axis.

The microsphere diameter is approximately 15 μm, and the sample size lies within the Mie scattering regime. Therefore, the model uses generalized Lorenz–Mie theory to calculate the axial optical force of the optical field. The axial motion of the microsphere is calculated using an overdamped model:

$$\gamma\, dz/dt = F_z(z,P) + F_g + F_b + F_{flow} + F_{thermal} + \xi(t) \tag{8}$$

Here, $\gamma$ denotes the viscous damping coefficient, $F_g$ and $F_b$ denote gravity and buoyancy, respectively, $F_{flow}$ denotes the force exerted by the background fluid, $F_{thermal}$ denotes the photothermal effect, and $\xi(t)$ denotes Brownian fluctuations. For an approximately spherical sample, the viscous damping coefficient is generally estimated as $\gamma = 6\pi\eta a$, where $\eta$ denotes the dynamic viscosity of the medium and a denotes the sample radius.

We performed numerical simulations to investigate the particle motion and optical force distributions. **Figure 3(a)** shows the focused optical field and the simulated particle trajectory. The optical intensity peaks near the focal point and gradually decreases in both the radial and axial directions. Under the resulting optical forces, the microparticle moves predominantly along the Z direction while remaining close to the optical axis, demonstrating simultaneous axial propulsion and transverse confinement.

The calculated optical-force distribution acting on the microparticle is detailed in **Figure 3(b)**. Under low-NA focusing, the axial gradient restoring force is insufficient to balance the forward scattering force, resulting in a net axial force along the direction of light propagation that continuously propels the microparticle. Meanwhile, the transverse gradient force remains directed toward the optical axis and provides lateral restoring confinement. The scattering and gradient forces therefore provide axial propulsion and transverse confinement, respectively, enabling stable particle transport along the optical axis through the light-sheet region.

**Figures 3(c)** and **3(d)** trace the particle displacement and optical-force variation, respectively. The displacement occurs predominantly along the Z direction, with only small variations in the X and Y directions, consistent with the combined axial propulsion and transverse confinement. As the microparticle approaches the focal region, the increased local optical intensity produces a stronger optical force; as it moves away from the focus, both the optical intensity and the corresponding force gradually decrease. These results demonstrate that the low-NA focused field enables continuous axial transport through the scattering force while the transverse gradient force confines the microparticle near the optical axis. The simulated directional motion is consistent with the experimental observations, providing theoretical support for scattering-force-driven active scanning and subsequent 3D fluorescence imaging.

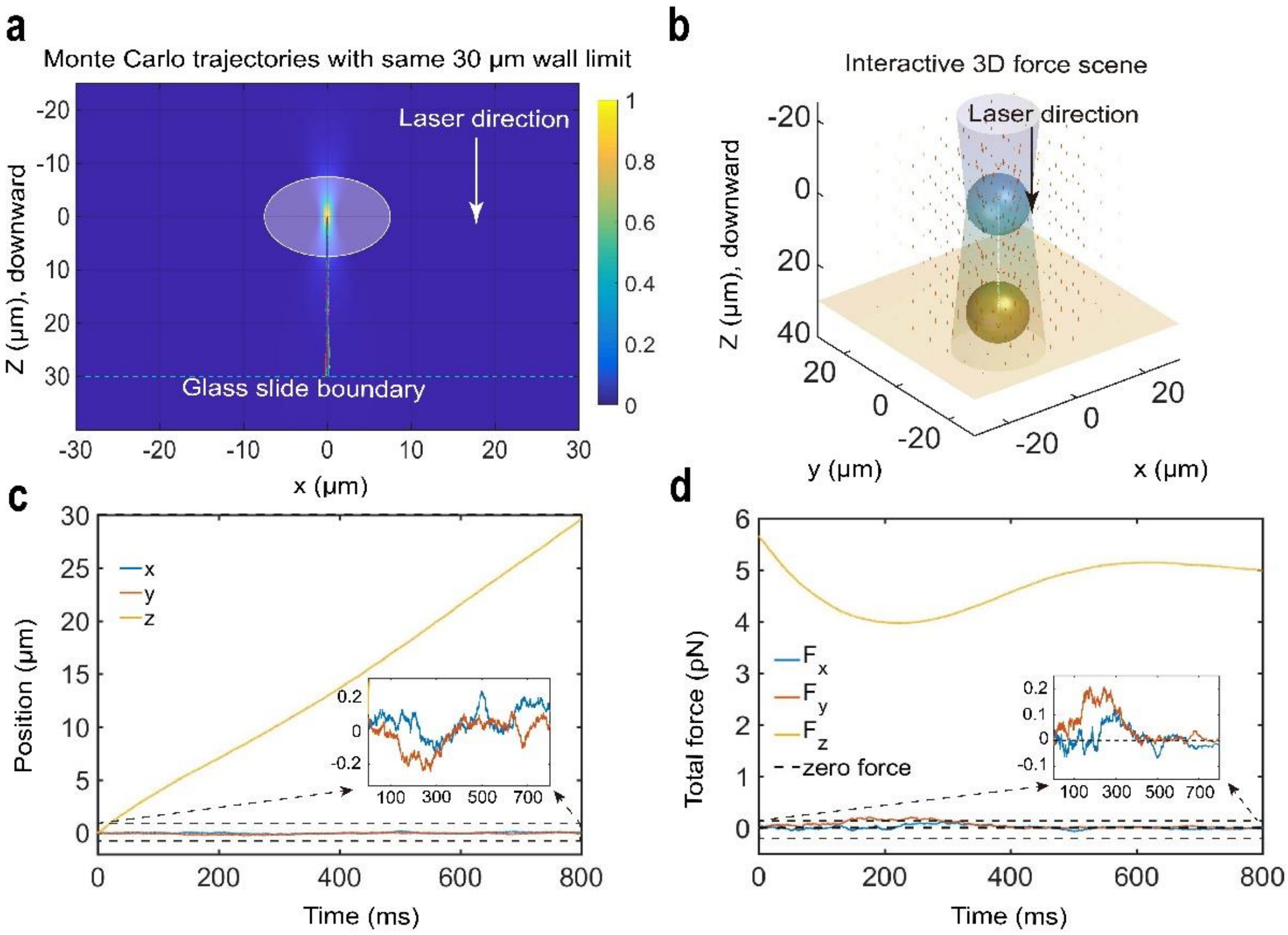

**Figure 3: Theoretical analysis of scattering-force-dominated optical propulsion. a,** Calculated trajectory of a 15-μm-diameter microsphere in a scattering-force-dominated optical field generated by focusing a 15 mW manipulation beam through an objective. **b,** Calculated directions of the optical forces acting on the microsphere over a 30 μm displacement, based on generalized Lorenz-Mie theory. **c,** Temporal variations in the simulated microsphere position along the X, Y, and Z axes. **d,** Time-dependent optical force components along the three axes. The axial force was approximately 5 pN, whereas the transverse force components remained below 0.2 pN.

### 3.2. Fluorescent polystyrene microspheres

Because continuous particle motion replaces mechanical stepwise scanning, accurately determining the axial position corresponding to each fluorescence image is essential for 3D reconstruction. To achieve this, a quadrant photodetector (QPD) is incorporated into the imaging system to obtain real-time particle positions from the detected scattering signal (Supplementary Note 1). In the synchronized acquisition scheme in **Figure 4(a),** a charge-coupled device (CCD) camera records fluorescence cross-sections as the scattering force propels the microparticle through the light sheet, while the QPD simultaneously measures the corresponding scattered-light signal.

The CCD images and QPD signals are synchronously acquired during the same particle motion and can therefore be correlated through their acquisition time. Using the pre-calibrated relationship between the QPD scattering signal and axial position, the scattering signal corresponding to each fluorescence image is converted into its actual Z-coordinate [42]. **Figure 4(b**) maps each tomographic image to its calibrated axial position, establishing a direct correspondence between fluorescence cross-sections and spatial coordinates for subsequent 3D reconstruction.

We evaluated the performance of active-scanning 3D light-sheet fluorescence imaging using 15-μm fluorescent polystyrene microparticles. Driven by the scattering force, a microparticle continuously passed through the light sheet while we synchronously acquired 150 fluorescence images and their corresponding axial positions. Representative fluorescence images at different time points are presented (Supplementary Figure 2). The entire dataset was acquired within 600 ms, with an acquisition time of approximately 4 ms per frame. By combining axial displacement calibration of the QPD using a nanometre-positioning stage with the position reconstruction algorithm (Supplementary Note 1), we achieved an axial sampling interval of approximately 0.1 μm, enabling rapid, finely resolved three-dimensional spatial sampling.

The 3D reconstruction was performed in MATLAB. Based on the axial coordinates obtained from the calibrated QPD signal, the preprocessed fluorescence sections were arranged according to their actual Z positions to generate volumetric data for 3D rendering (**Figure 4(c)**). The reconstructed microparticle exhibits a well-defined spherical morphology with a measured diameter of approximately 15.5 μm, in good agreement with its nominal diameter of 15 μm and corresponding to a relative error below 5%. These results indicate that the method can accurately recover the 3D morphology and dimensions of the sample.

To further evaluate the surface characteristics of the reconstruction, isosurface analysis was performed on the 3D fluorescence intensity distribution. The isosurface in **Figure 4(d)** is continuous and retains a regular spherical morphology, while local variations reflect differences in surface fluorescence intensity and texture. The reconstructed geometry is consistent with the expected spherical shape and smooth boundary of the polystyrene microsphere. These results demonstrate that light-sheet fluorescence imaging combining QPD-based position measurement with continuous active scanning enables rapid and accurate 3D reconstruction. This strategy provides a mechanical-scan-free active optical manipulation approach for 3D fluorescence imaging with a high image acquisition rate [37].

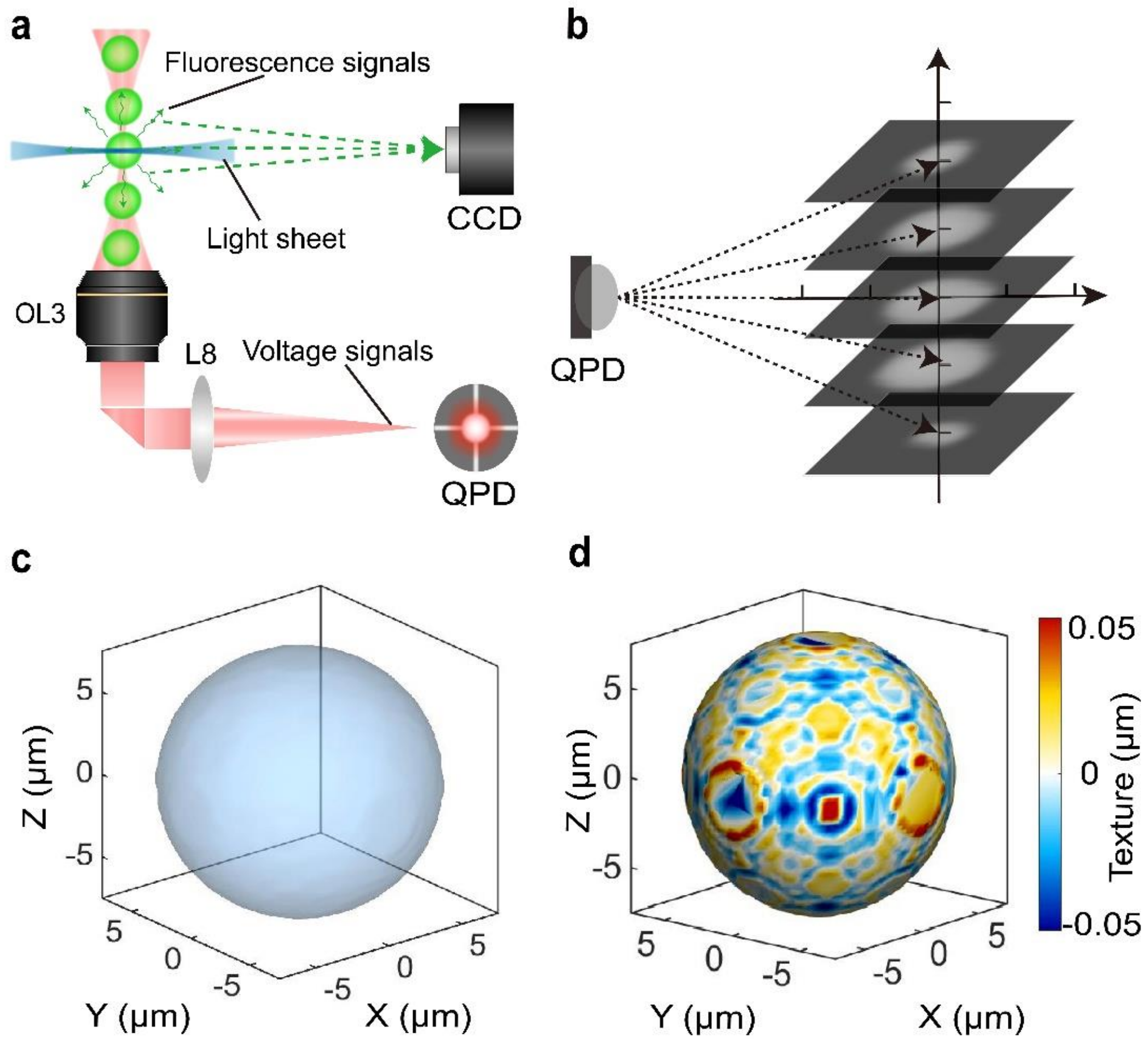

**Figure 4: 3D reconstruction of a 15-μm-diameter fluorescent polystyrene microsphere. a,** Light-sheet fluorescence signals recorded by the CCD camera and QPD voltage signals generated by the scattered manipulation beam as the microsphere was driven along the direction of light propagation. **b,** Axial coordinates assigned to 150 fluorescence images using QPD-based position calibration. **c,** 3D rendering of the reconstructed fluorescent polystyrene microsphere. **d,** Isosurface analysis of the reconstructed microsphere, showing surface topographical variations below 0.05 μm.

### 3.3. Fluorescent sodium alginate microparticles and agarose hydrogel magnetic beads

To further evaluate the imaging capability of the method for different spatial structures, fluorescent sodium alginate microparticles and agarose hydrogel magnetic beads, both approximately 15 μm in diameter, were selected as model samples. The morphology and geometric features of the different microsphere samples are shown in optical micrographs acquired under white LED illumination (Supplementary Figure 3). The two soft-matter samples exhibit distinct fluorescence and structural distributions. Fluorescence in the sodium alginate microparticles is primarily distributed in the outer layer, providing a model for evaluating surface morphology reconstruction, whereas the agarose hydrogel magnetic beads consist of a fluorescent hydrogel matrix containing non-fluorescent magnetic particles, allowing the detection of internal heterogeneous structures to be assessed.

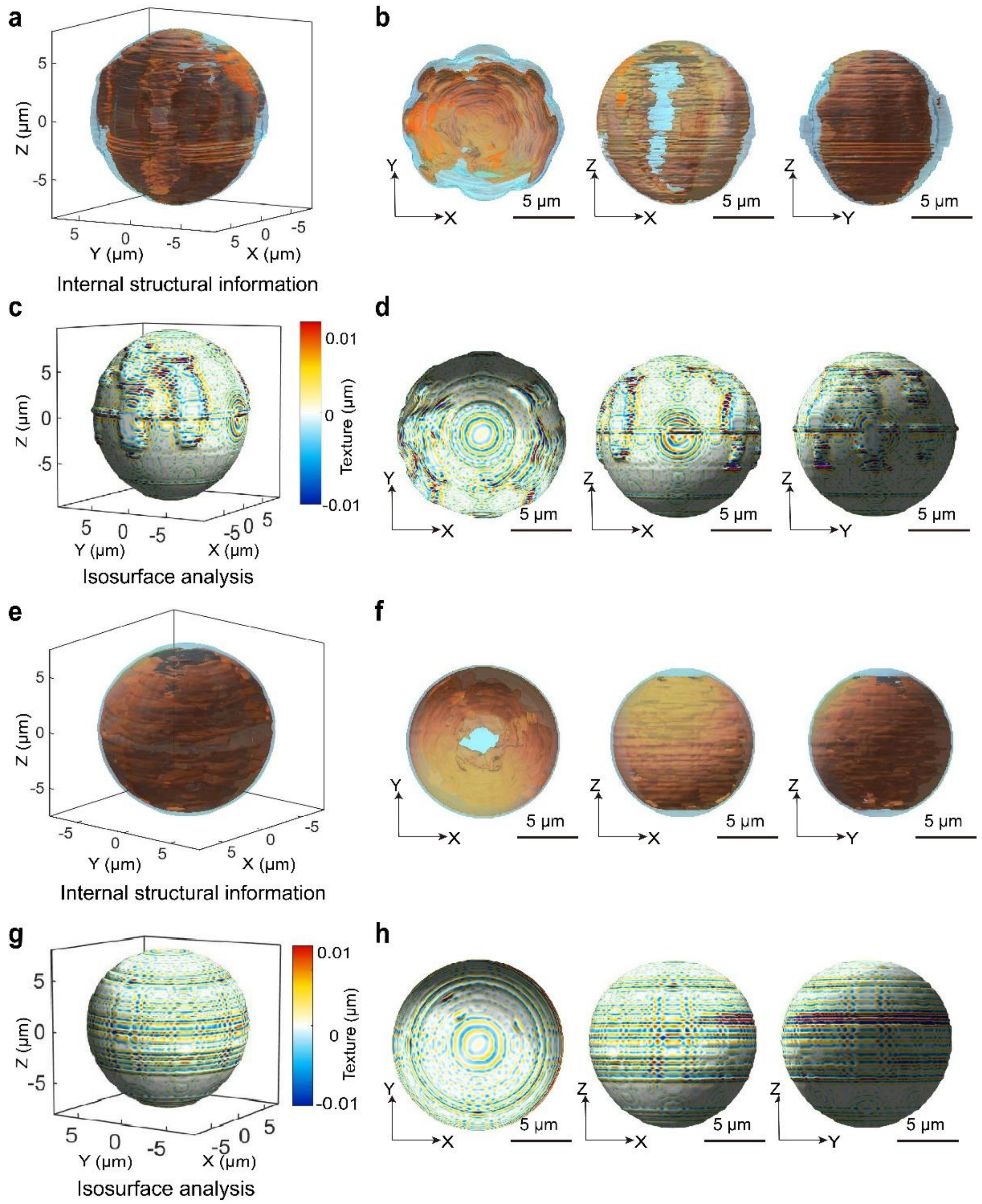

**Figure 5: 3D reconstruction of fluorescent sodium alginate microspheres and fluorescent agarose hydrogel magnetic beads. a-d,** Fluorescent sodium alginate microsphere; **e-h,** fluorescent agarose hydrogel magnetic bead. **a, e,** 3D volume renderings of the reconstructed

samples. **b, f,** XY, XZ, and YZ views of the 3D reconstructions. **c, g,** Isosurface analyses of the 3D reconstructions. **d, h,** Corresponding XY, XZ, and YZ views of the isosurface analyses.

**Figures 5(a)-(d)** document the reconstructed three-dimensional morphology of fluorescent sodium alginate microspheres. The reconstruction exhibits an approximately spherical shape, with pronounced local surface undulations and a nonuniform fluorescence distribution. Across different viewing angles, spatial variations in local morphology become apparent; isosurface analysis further emphasizes the surface texture and its distribution.

**Figures 5(e)-(h)** characterize the three-dimensional structure of agarose hydrogel magnetic beads. Compared with the sodium alginate microspheres, the reconstructed beads exhibit a more regular overall contour but distinct variations in internal fluorescence distribution, with localized fluorescence-deficient regions corresponding to nonfluorescent magnetic particles embedded in the hydrogel. Multiplanar reconstructions localize these regions within the fluorescent matrix, and isosurface analysis helps delineate internal heterogeneity from fluorescence-intensity variations.

Together, these results indicate that the method can recover nonuniform surface morphology and identify internal nonfluorescent components within a fluorescent matrix, highlighting its versatility across distinct microscale structures. Scattering-force-driven active-scanning light-sheet fluorescence imaging therefore provides a new approach for 3D structural characterization of soft matter, cell-like models, and complex microscale samples [43].

## 4. Discussion

Our approach combines scattering-force-driven continuous motion with real-time position measurement, offering notable advantages in reducing the axial sampling interval, increasing the image acquisition rate, and lowering the optical power required for manipulation without relying on complex scanning devices. **Table 1** compares the present method with different 3D microscopy techniques. Recent advances in integrated optofluidics offer new opportunities for extending this approach. Previous studies have integrated microchannels, light-sheet illumination, and on-chip optical components to achieve continuous 3D imaging of suspended cells [44,45]. Combining these technologies with our method could enable continuous sample transport and lateral positioning within microchannels, followed by scattering-force-driven axial scanning, extending single-particle 3D imaging to automated, high-throughput spatial information acquisition. Beyond high-throughput 3D imaging, combining this approach with dual-comb spectroscopy could provide complementary spectral information, further enabling multimodal characterization of microscale samples [46,47].

The present method has mainly been demonstrated using microscale model samples. Variations in the size, refractive index, and irregular morphology of biological cells may alter their optical-force response and position-detection signals, while flow perturbations and sample rotation may introduce additional uncertainty in spatial registration. Developing sample-specific optical-force models and optimizing dynamic position calibration and flow control could improve motion stability and reconstruction reliability for complex samples [11]. A smaller axial sampling interval does not necessarily imply higher optical resolution, and further improvements in axial resolution require consideration of light-sheet thickness, position-detection errors, and image reconstruction methods [48]. For live-cell imaging, photobleaching and cell viability assays are also needed to assess the effects of optical illumination. These improvements could further extend the method to 3D phenotypic analysis of suspended cells, characterization of microgel carriers, and dynamic imaging of microscale structures.

## 5. Conclusion

We propose a scattering-force-driven active optical manipulation method for 3D imaging that combines continuous directional particle transport with real-time position measurement to enable mechanical-scan-free 3D light-sheet fluorescence imaging. The method achieves an axial sampling interval of 0.1 μm, acquires fluorescence images at 250 Hz, and enables 3D reconstruction of microscale samples with distinct structural features, while reducing the required manipulation power by approximately 70%. By extending optical manipulation from stable trapping to active spatial sampling, this approach provides a new strategy for rapid, low-power 3D imaging and dynamic characterization of complex microscale samples.

**Table 1. Comparison of different 3D microscopic techniques**

| Parameters | This work | Ref. [14] | Ref. [29] | Ref. [19] |
|---|---|---|---|---|
| **Scanning mode** | Axial scanning | Axial scanning | Axial scanning | Rotational scanning |
| **Driving mechanism** | Scattering-force | HOT | Mechanical displacement | MCF-OCR |
| **Scanning frequency** | 250 Hz | 10 Hz | 62.5 Hz | 7.1 Hz |
| **Imaging method** | Photoluminescence | Structured illumination | Photoluminescence | Optical projection tomography |
| **Step size** | 100 nm | 160 nm | 250 nm | 0.72° |
| **Laser power** | 15 mW | 50 mW | / | 80 mW |
| **Position fluctuation** | Effectively avoided | 50 nm | / | 1.49° |